# USE OF ANALOGIES IN SCIENCE EDUCATION, A SYSTEMATIC MAPPING STUDY

Hernandez Pedro and Espitia Edinson

Master of Education, Castle University, university of Cordoba, Colombia

*ABSTRACT*

*This systematic mapping study consisted of tracking the scientific literature that addresses the issue of analogies as a didactic strategy in science teaching. An analogy can be understood as comparing an existing knowledge with a new knowledge to achieve a better understanding of the new knowledge as a result of the comparison of similarities; or in other words, use students' own concepts to introduce new concepts using comparisons between the two. The purpose of this study was to identify, analyze, synthesize and evaluate research works that touched on this topic, with this, to have knowledge about the models of uses of analogies, most used didactic strategies, research methodologies in this field and how to evaluate the learning effectiveness of working with analogies. The methodology that was used is the systematic mapping study; Five questions were posed that guided the information tracking process. Later, the electronic documents in English for the last twenty years were traced in five databases related to the educational field. Finally, it is concluded by responding to the purpose of the study where it is evident that, broadly speaking, the research methodologies in this field are quantitative as well as qualitative, to implement analogies, resources such as images, illustrations, textual indications and audiovisual aids are used, it is usually evaluated the effectiveness of using analogies with multiple choice tests, oral tests of creating analogies by students.*
.

## 1. INTRODUCTION

A quality education indicates aspects such as infrastructure, qualified personnel and excellent pedagogical processes that promote active learning of students. Thus, improving educational quality, one way or another, leads to an emphasis on pedagogical processes and this represents a special focus on the teaching tools that are implemented in classrooms in order to strengthen pedagogical processes. Analogies over time have been an excellent didactic tool that allows improving the teaching-learning processes.

Harriso y Treagust [1], argue that analogies have long formed tools of discovery in science, and are often used as explanatory devices within the classroom. However, research has shown that analogies can elicit alternative conceptions because some students view the analog differently than the teacher. This position is classic in this field and is the basis for later studies on this particular issue. It should be understood that "analogies are comparisons between relationships from a familiar domain (base) to an unknown domain (target). Due to this, they can be important ways for teaching scientific concepts and at the same time, the analogies drawn by teachers can show their conception, their beliefs and mainly the kn they mobilize when drawing them " [2, p. 6]

In other words, "an analogy can be defined as a comparison between the similarities between two different domains, one unknown, or little known, which can be called destination domain and last known destination, for which it is called the analogous domain" [3, p. 5]. In this regard, Guerra [4], confirms that an analogy is a comparison of the similarities of two concepts, the familiar concept that is called the analogous and unfamiliar white. Both the analog and the target have characteristics of their own, (also called attributes). An analogy can be drawn between them. A systematic comparison, orally or visually, between the characteristics of the analog and the target is called a mapping.

There "Analogy is considered a useful way to help students visualize abstract concepts and assimilate new insights from an existing cognitive structure" [5, p. 5]. For this, there are various resources and models to implement analogies in classrooms, Kepceoğlu and Karadeniz [6] whos show that analogies are classified with respect to their analogical relationship, presentation format, abstraction, position of analog with respect to the objective and wealth level.

In this sense, "The creation of analogies are a powerful tool for explanation, as well as a fundamental mechanism to facilitate the individual's construction of knowl, [7 p.5], however these can be "two-edged swords Because the knowledge they generate is often accompanied by alternative conceptions, when people receive analogies, they use their past knowl, experiences, and preferences to interpret the analogy to harmonize with their current thinkin [1, p. 3].

For his part Guerra [4], suggests that analogies as pedagogical tools have strengths and limitations; analogies are clearly not a magical resource to promote understanding of science or its concepts, but their potential can be capitalized if they are used wisely.

However, Harrison y De Jong [8] recommend the use of multiple analogies and insist that teachers always show where the analogy breaks and carefully negotiate conceptual results. Indeed, "the challenge of science education is then to help service teachers to use analogies when teaching science concepts" [7 p. 3].

In this sense, Harrison y Treagust [1] affirm that science classrooms are a common environment in which analogies are used to improve for learning concepts; therefore, improving the way analogies are used in science education, this has had important consequences for improving teaching and learning processes.

De Rosa, Pimentel, y Terrazzan [3] consider that, in the context of teaching and learning subjects in the field of natural sciences, analogies favor students in understanding a scientific domain that is unknown for them, based on a family domain, by comparing common and uncommon attributes and relationships between both domains.

Indeed, "analogical reasoning is known as an important mental process for solving problems of everyday life, specialized problems in mathematics, physics and chemistry, in a creative way. When a new problem must be solved, people remember similar problems that they have solved before and adjust the solutions they used at the time according to the nature and characteristics of new problems to solve them" [9 p. 9].

In accordance with this, a research study was proposed on the use of analogies for science teaching, where the development of contextualized analogies to the rural environment will be addressed, following guidelines of pre-established models, for the development of analogies in said context. In this way, the present work shows the initial phase of this research aimed at developing a systematic mapping study that allows a comprehensive overview of the use of analogies in the teaching of science, methods, didactic resources that are used, techniques and instruments used in existing research, carried out on this specific area of knowledge.

This document is organized as follows: First, the definition of the research questions; second, conducting the search; in the third moment, the classification of documents; in fourth place, the extraction of data and finally in fifth place, the preparation of the report and the publication.

## 2. SEARCH METHOD

At the beginning of the process of carrying out the systematic mapping, it was necessary to take into account a series of steps, which will allow to be more successful with the search and tracking of information. Understanding that "a systematic review of the literature is a means of identification, evaluation and interpretation of all available research that is relevant to a specific research question, thematic area or phenomenon of interest" [10, p. 11].

In other words, systematic mapping is a technique that allows to account for the knowledge that exists on a specific topic, in this way, to be certain of scientific advances, as well as the lines of research that occur in a certain field of science. knowledge.

Now, the term systematic mapping is often associated or confused with another type of scientific research such as systematic review, however, they are two different concepts that respond to different purposes and procedures, on this Katy L. James, Nicola P, Neal R. and Haddaway [11] clarify that systematic mapping does not attempt to answer a specific question like systematic reviews, but rather collects, describes and catalogs the available evidence.

Therefore, the steps that will be followed for the elaboration of this systematic mapping are those proposed by Kitchenham [10].

*2.1. Definition of research questions.*
*2.2. Conducting search.*
*2.3. The classification of the works.*
*2.4. Data extraction.*
*2.5. Preparation of report and publication.*

### 2.1. Definition of Research Questions

In order to track information that responds to the topic of using analogies for school teaching in rural contexts, five research questions were defined:

RQ 1. What evidence indicates the use of analogies to facilitate science learning?
RQ 2. What types of models for the use of analogies are used for the development of science?
RQ 3. What type of research methodology has been used to develop studies that use analogies for science teaching?
RQ 4. What type of teaching resources have been used in investigations that use analogies to teach ashes?
RQ 5. What methods have been used to assess learning in studies that use analogies for science teaching?

With these five questions, an attempt was made to cover the subject matter in question in a very complete way and thereby give an account of how research is developing in this field of knowledge, in that order of ideas;

RQ 1: I seek to track information that indicates how analogies have been used as a didactic tool for teaching science, to facilitate teaching-learning processes in classrooms;

RQ 2: I inquire about the way in which analogies have been implemented, that is, criteria, processes and / or steps to use analogies in a correct and effective way, within the classroom;

RQ 3: aimed to identify which research methods have been used in this field, with this we refer to the type of study, design and techniques most used;

RQ 4: I seek to recognize what types of didactic resources are useful to implement analogies (comics, images, among others), and

RQ 5: gave an account of the way they have been used to evaluate the effectiveness of analogies as a didactic tool in the teaching-learning process.

## 2.2. Conducting Search

To search for the required information, a procedure was carrie out where the research categories were reviewed, then the search strings were elaborated and as a third step, the data bases where the literature searches will be carried out were defined.

This, a review of the categories immersed in the investigation was made, a necessary process to start the development of search steps, that is; It is necessary to be clear about the categories that are going to be worked on in the study, since these are the basis of support for the next step, which will result in the definition of the search strings.

This is important as "defining search parameters implies the selection of search terms, combining them with the help of search operators, and using the appropriate search fields"[12 p. 13], from this it can be inferred that, to build a search chain, it is necessary to be certain of the categories to be investigated and thus use the appropriate terms.

Now to design an appropriate search string. Kitchenham [10], presents a guideline on the definition of these strings to reduce the desired search bias. This, we select the keywords related to the research questions to compose the research chain.

This aspect is important to give credibility and confidence in the investigative process by reducing biases in this process. Consequently, when building a search string, it is important to "define and use search parameters that have a significant impact on search results, so researchers often have to experiment with combinations of search terms and operators [12 p. 14]."

Vom Brocke, Simons, Riemer, Niehaves, Plattfaut y Cleven 2015 [12] when using several search operators to improve the results in the use of search terms, for example, combining them with Boolean operators in search phrases [for example, AND, OR]. are while the exact use and interpretation of the search phrases can be determined with the help of parentheses, the precision of the search can be influenced by quotation marks.

Likewise, Boell and Cecez-Kecmano [13] argue that, to define search strings, the combination of terms using 'OR' indicates that any of the terms used is sufficient to retrieve a document. An important use of 'OR' is to include synonyms in a search. In contrast, 'AND' is used to restrict a search to documents that show different conditions at the same time. Its main use is to join different terms. These terms can represent different concepts that should be show the present at the same time.

Taking into account the previous approaches regarding how the search strings should be elaborated, three search strings were defined, where it is intended to give answers to the research questions and the following data bases were chosen: jstor, scopus, taylor & francis, proquest and science Direct.

Table 1:

| PREGUNTAS | STRING |
|---|---|
| 1. What evidence indicates the use of Analogies to facilitate science learning of primary school students in rural educational contexts?<br>2. What kind of models of use of analogies is used for the science teaching in rural educational contexts?<br>3. ¿What type of research methodology has been used for the development of studies that use analogies for science teaching in rural educational contexts? | ("Use of Analogies" OR "analogies" OR "analogy") AND ("scientific education" OR "Science Teaching" OR "science learning") AND ("primary education" OR "elementary school" OR "primary school") AND ("rural area" OR "rural education") |
| | ("Use of Analogies" OR "analogies" OR "model analog") AND ("Science Teaching" OR "science learning") AND ("primary education" OR "elementary school" OR "concept teaching") |
| 4. ¿What kind of teaching resources have been used in studies that uses analogies for science teaching? | ("didactic resources" OR "resources" OR "pedagogical resources") AND ("Use of Analogies" OR "analogies" OR "analogy") AND ("scientific education" OR "Science Teaching" OR "science learning") AND ("primary education" OR "elementary school" OR "primary school") AND ("rural area" OR "rural education") |
| 5. What methods have been used to evaluate learning in studies that uses analogies for science teaching? | ("instrument of measurement" OR "evaluation" OR "assessment" OR "evaluation instrument" OR "assessment instrument" OR "evaluation method") AND ("Use of Analogies" OR "analogies" OR "analogy") AND ("scientific education" OR "Science Teaching" OR "science learning") AND ("primary education" OR "elementary school" OR "primary school") AND ("rural area" OR "rural education") |

Table 2:

| # | STRING | TOTAL |
|---|---|---|
| **STRING 1** | ("Use of Analogies" OR "analogies" OR "analogy") AND ("scientific education" OR "Science Teaching" OR "science learning") AND ("primary education" OR "elementary school" OR "primary school") AND ("rural area" OR "rural education") | 1.834 |
| **STRING 2** | ("didactic resources" OR "resources" OR "pedagogical resources") AND ("Use of Analogies" OR "analogies" OR "analogy") AND ("scientific education" OR "Science Teaching" OR "science learning") AND ("primary education" OR "elementary school" OR "primary school") AND ("rural area" OR "rural education") | 1.089 |
| **STRING 3** | ("instrument of measurement" OR "evaluation" OR "assessment" OR "evaluation instrument" OR "assessment instrument" OR "evaluation method") AND ("Use of Analogies" OR "analogies" OR "analogy") AND ("scientific education" OR "Science Teaching" OR "science learning") AND ("primary education" OR "elementary school" OR "primary school") AND ("rural area" OR "rural education") | 2.286 |
| **STRING 4** | ("Use of Analogies" OR "analogies" OR "model analog") AND ("Science Teaching" OR "science learning") AND ("primary education" OR "elementary school" OR "concept teaching") | 7.661 |

The information tracking only included digital databases; with a total of five databases, of importance in relation to educational issues, which were already mentioned above. Four different strings were elaborated, in table 2, the total results of documents found in each of the different strings are shown, thus, in string 1 1,834 results were found, in string 2 a total of 1,089 and in string 3 2,286 results were found and in string 4 7,661 documents were tracked.

For the following stages of this study, string 4 will be taken into account, since they present a greater number of tracked documents, in order to address each of the research questions in a comprehensive and complete way, and this will be chosen trace string. Now, this information tracking process began in February 2020 and lasted until the middle of June 2020.

### 2.3. The Classification of the Works

In order to classify and carry out a detailed review of the tracked literature, it was necessary to have defined criteria of inclusion and exclusion in this way to be able to emphasize the documents that are relevant to our work, thus avoiding futile efforts in excavtive reviews of literature that do not fit under the research objectives, and do not contribute to answering the research questions already posed.

Similarly, having clear and well-defined criteria to exclude and include information will allow to reduce biases in the investigation process, which indicates that, as long as these criteria are structured and the relevant steps are followed, it will be carried out. more reliable research.

Table 3:

| Inclusion criteria | Exclusion criteria |
|---|---|
| 1. Primary studies that show the use of analogies in the teaching and learning processes of science.<br>2. Studies that specify models for the use of analogies in education.<br>3. Articles that detail the research methodology used in educational interventions that use analogies<br>4. Research articles that show the different resources used to implement analogies in the educational field. | 1. Articles that do not address analogies, although they do inquire about another category of the study.<br>2. Articles that are not in English since the universality of this language globalizes research.<br>3. Articles that do not present full text and are not available in electronic format.<br>4. Technical reports, documents that are available in the form of abstracts or presentations, as well as secondary literature reviews.<br>5. Duplicate publications, it is considered the most recent in redundant studies of the same authorship.<br>6. Articles that are not research products (systematic reviews, interpretive studies from other contexts).<br>7. all documents that are more than 20 years old, that is, those that were published before the year 2000. |

### 2.4. Data Extraction

At this stage, the documents were read completely and in detail to answer the questions that were raised at the beginning and finally to obtain the evidence that supports this mapping study.

Table 4:

| STRING | DATABASES | DOCUMENTS INCLUDED | DOCUMENTS EXCLUDED | TOTAL |
|---|---|---|---|---|
| ("Use of Analogies" OR "analogies" OR "model analog") AND ("Science Teaching" OR "science learning") AND ("primary education" OR "elementary school" OR "concept teaching") | **ProQuest One Academic** | 22 | 6.504 | 6.526 |
| | **Jstor** | 4 | 596 | 600 |
| | **Taylor & Francis** | 1 | 263 | 264 |
| | **Science-Direct** | 0 | 268 | 268 |
| | **Scopus** | 1 | 2 | 3 |

Table 5:

| | DATABASES | DOCUMENTS INCLUDED |
|---|---|---|
| ("Use of Analogies" OR "analogies" OR "model analog") AND ("Science Teaching" OR "science learning") AND ("primary education" OR "elementary school" OR "concept teaching") | **ProQuest One Academic** | 22 |
| | **Jstor** | 4 |
| | **Taylor & Francis** | 1 |
| | **Scopus** | 1 |
| **TOTAL DOCUMENTS INCLUDED** | **28** | |

### 2.4.1. Results

To have greater clarity about what was obtained in the search, three important aspects about the literature searched will be mentioned; in this way, in the first place, the means of diffusion, the type of research method and the population under study.

It was found that 39.29% (11 documents) corresponded to theses to aspire to the doctoral degree published in journals, 7.14% (2 documents) went to theses to aspire to the master's degree published in journals, 25% (7 documents) corresponded to research articles published in journals, 14.28% (4 documents) were book chapters and finally 14.28% (4 documents) were published in conferences.

On the other hand, in terms of methodological research, 17.85% (5 documents) had a qualitative approach of case study and action research, 17.85% (5 documents) corresponded to the non-experimental quantitative approach, 14.28% (4 documents) They were of mixed methodology, 28.57% (8 documents) were quantitative quasi-experimental and pre-experimental with

94      Computer Science & Information Technology (CS & IT)

intervention, and 21.42% (6 documents) corresponded to a quantitative experimental approach of the Solomon type and pretest-posttest with control group.

Regarding the study population, 25% (7 documents) were from primary school, 57.14% (16 documents) corresponded to secondary school, 14.28% (4 documents) were undergraduate and 3.57% (1 document) were from a population open, that is, they included a different population.

### 2.4.2. Answers to Research Questions

*Evidence the use of analogies to facilitate science learning*

This question inquires about those investigations that show, how analogies in practice can be considered as a powerful tool when teaching concepts in science, an attempt was made to trace all the investigations that included analogies as a didactic strategy to impart knowledges.

Consequently, all the documents that were included for the present study provide evidence of the value of analogies in teaching scientific concepts. It should be noted that 32. 14% (9 documents) of the investigations found study how analogies are used in science textbooks, and the way in which these analogies, present in teaching guide books, can improve the learning of science students.

67.85% (19 documents) are studies that investigate how analogies as a basic didactic strategy in science classes directly influence student learning and contribute to achie meaning ful learn in science. Everything found and presented in this study shows the great importance of analogies in the educational field, and especially in the school teach of scientific concepts, however, it must be borne in mind that, although this document only addresses studies published in the last 20 years, the topic of analogies in the research field as a vital didactic tool in the teach-learn processes, has been studied for several decades behind those contemplated by this study.

*Models for the use of analogies in the teaching of science*

thats all the research retrieved, only twelve studies were identified where models for the use of analogies in science were mentioned, thus providing an answer to this question.
Rule A, Baldwin S y Furletti [14], The authors showed a way to combine form and function with analogy is through the form and function analogies of boxes of objects. This technique has been used successfully in teaching, to use the object boxes feature form to teach animal adaptations to third graders.

Bennett-Clarke, B Ciana. [15], They take the model of (Duit, 1991; Wittrock and Alesandrini, 1990; Wong, 1993a, b) On the analogies built by the teacher and the analogies that are generated by the student self-generated analogies. Thus, self-generated analogies are meta-cognitive strategies, founded on Generative Learning Theory. They are known to facilitate the understanding of abstract ideas by pointing out similarities familiar to the student in the real world, promote self-generation of meaning and autonomy in learning, and awaken students' interest and motivation to learn.

M Sota [16], They studied contrasting analogies: situations that are not analogous to the concept or principle for which a false idea is carried out. Given the well-documented effects of non-examples on conceptual learning and research on contrasting analogies in problem solving, one might expect that contrasting analogies will have a positive effect on conceptual change.
L Asay [17], Cite models for teaching analogies, highlight the objectives of each, claim that the few existing models for instruction with analogies have often not been quantitatively examined.

The model with analogies is one of the models frequently cited in the variety of investigations on the subject. The model outlines the steps for instruction, including the step of explicitly mapping characteristics from source to destination.

Second model, teaching with analogies Provides principles for evaluate instructional analogies and steps for using analogies, but relied on analysis in science textbooks, rather than pedagogical principles or research on effectiveness.

A third model, the Focus, Action, Reflect or FAR teaching model emerged from Australia, with the use of TWA model teachers. This model includes planning the use of analogies and an evaluation of the process, but again focuses on what teachers do to prepare.

L Atkins [18], developed a count of analogies generated as categorization statements using the events in this classroom investigation from categorization.

B Gary [19], I implement the teaching model that I call the Pre-Analog Step: all to teaching method to improve students' familiarity with important analog features in an analog source before introducing an analogy to understand a target domain.

L Newton [20], Addressed the position of the analogy in the instructional text that can vary. You could introduce yourself as an analog advance organizer, at the beginning of the instructional unit. It could be integrated into instruction at a point where new and / or more difficult information is introduced, what Curtis calls a built-in trigger. As such, it can also help clarify what has happened before and lead to new but related material. Finally, the analogy could occur at the end of the instructional text, to act as a vehicle and synthesize what had happened before and conclude the topic, described as a later synthesizer.

So far, analogies have been mentioned in an undifferentiated way. They can be identified as: Simple: analog statements without elaboration; Enriched: analog statements that include the basis of the analogy; and extended: analog statements that can be applied to various topics and used in various contexts.

K Clement y K Yanowitz [21], show a situation model that includes not only text entities, but also relationships between text entities such as temporal and causal relationships. There, the situation model includes explanations of why text events occur. A well-developed situation model requires an understand of the causal mechanisms explicitly described in the text, as well as inferences about the causal mechanisms. The situation models of a text are. they build on the basis of both the text itself and prior knowledge, was Schémas, general factual knowledge, or knowledge specifically related to the current text.

Madera S [22], mentions three teaching models, have also been identified. WWA is a derived model that adapts to the ideas of both GMAT and TWA. The Working with Analogies model should appeal to teachers as well as students because of its fit with current fashionable constructivist approaches to the teaching and learning of physics.

Murat G [23], I use the student-centered model of analogies. A study was carried out with 6th grade primary school students. Where the students were asked to form their own analogies as a result of the study in which the student-centered analogy technique was used.

Dundar Y [24], It is a study on the analogies presented in the high school physics text books. This model is also called the "raisin cake model" as it looks like raisin cakes This is explained in the structure analogy by comparing the structure of an atom to a raisin cake. According to previous research, multifunctional analogies have been used primarily in chemistry text books.

Musa D [25] They mention that, for analogies to become effective teaching tools in biology textbooks, they should be created based on guides such as Teaching with Analogies and Focus-Action-Reflection ( FAR; Treagust et al., and its limitations should be explicitly described and systematically presented.

*Research methodologies in studies on analogies.*

After the trace of the literature, it was found that the investigations on analogies rotate in the three methodological directions, that is, works of methods were obtained. qualitative, quantitative and mixed studies were also found. It was observed that 17.85% (5 documents) were of a qualitative approach of case study and action research, 17.85% (5 documents) corresponded to the non-experimental quantitative approach, 14.28% (4 documents) were of mixed methodology, 28.57% (8 documents) were quantitative quasi-experimental and pre-experimental with intervention, and 21.42% (6 documents) corresponded to a quantitative experimental approach of the Solomon type and pretest-posttest with control group.

*Didactic resources used in investigations that implement analogies.*

This question inquires about those teaching tools or ways of presenting analogies when implementing them in teaching processes. In this way, it was found that the teaching resources used to implement analogies are: verbal instructions by the teacher, videos, images or illustrations [26], [27], [28], [29], [30], [ 31], [32], [33], [34], [35], [36], [37], [38], [39], [40], [41], [42], [43] .

It should be mentioned that in the different models that were found, both instructions and images are used in the same analogy, although only the instructions of the teacher or expert can be used to explain the analogy and the relationships to be related.

*Methods used to assess post-analog learning learning*

When analogy is used as a didactic tool or strategy for better learning, the literature consultation clearly evidenced that there are some reforms to evaluate these processes, in this way to verify how analogies impact the teaching-learning processes.

In the documents tracked, it was found that tests with multiple selection were carried out with several evaluation items, as in the study of. Nita A. [27] where after the text study, the participants returned to complete the questionnaire of nine items, designed to retest your subject interest, perceived knowledge, and self-efficacy in teaching all three concepts.

In addition, the participants were administering text measures and were asked to classify the texts they read with respect, the participants were asked to complete the text by measure the classification of the three versions, with respect to how interesting it was, how useful it was the text to understand the concept and how helpful the text was from the perspective of explaining the concept to fifth graders.

Another type of evaluation that was found was the one proposed by Rule A, Baldwin S. y chell [38]  where they mapped analogies, thought about alternative manufactured articles and created new analogies. Smith C. (2007) shows that they participated in individual interviews, before and after the curriculum unit received written pre and post exam assessments from their teacher.

On the other hand, Clement K y Yanowitz K [21] designed three tasks to assess what information the participants believed was central to the target domain. Object Task

The first task, revealed the participants' beliefs about the importance of individual objects in their situation model of the objective passage, and circled 4 objects that you consider most important to the story. "The relationship task Between Objects and the Relationships Between Events Task for the final task, participants had to do similar work to the previous one, except that instead of forming lower-order relationships, they were based on a given list of objects.

Similarly, evaluations such as those by Chuang M y Hsiao-Ching [41] stand out, in which the participants were interviewed and recorded before, a week after and 7 weeks after learning. Bennett-Clarke C. in this study the researcher made a five-minute summary of the lesson generating a discussion, asking the students to identify and briefly summarize the analogy that was used in that lesson of the day the students were assigned approximately three to five minutes by the end of the lesson. The summary for recording the analogy in your analogy journals and completing the journal entry page. In summary, each page of the magazine asked them to identify the analogy that their teacher presented, brief explain the relationship between analogy and concept, and to complete the four self-report items for situational interest, participating students should complete their journals for a project grade over the next two weeks of the course, students were asked to generate their own analogies. Independently in class and out of class based on the concepts covered. Analogies that students were expected to generate during the intervention period mainly simple surface analogies.

Finally, another evaluation strategy to highlight is that proposed by Asay L. [17] here the subjects took a previous test before the instruction, which was used to assign them to a previous level of knowledge about electrical circuits for the analysis of any effect differential. After the instructional modules, students took a posttest on electrical circuits. Two weeks later, they took a delayed posttest. Three different forms of a 20-question multiple-choice test on electrical circuits were used for the pretest, posttest, and delayed posttest.

A bank of 60 multiple, matching and true false options. Questions about electrical circuits were developed. The exam questions were prepared based on a set of item specifications for the components of the electrical circuits topic.

Although most studies in one way or another show how to evaluate procedures that use analogies, those that are different and present different characteristics are highlighted, in reference to the rest of the studies that fit within the mentioned evaluations.

## 3. CONCLUSIONS

With the detailed review of the twenty-eight documents that were considered relevant within the search, the purpose of this study is answered, which was to identify, analyze, synthesize and evaluate research works that touched on this topic, thereby, having knowledge about the models for the use of analogies, most commonly used teaching strategies, research methodologies in this field and how to evaluate the learning effectiveness of working with analogies.

In this way, it was found that there are several teaching models with analogies (more than three models highlighted in the literature, as shown above where this question was answered in previous pages), a variety of didactic resources that are useful to implement analogies such as illustrations with images, videos, verbal instructions among others. Regarding the way to evaluate the effectiveness of analogies, it was evidenced that written and verbal tests are applied, the creation of new analogies by the students as well as verbal reasoning about the conceptual change.

These findings can contribute to the decision-making processes on methodologies, indicators, strategies, techniques and instruments that will be used in conducting development-oriented research in the teaching-learning of science and / or analogies as a way of teaching. This study leaves certainty about the fundamental role that analogies have had over time, especially in the last twenty years, as a fundamental strategy that facilitates teaching-learning in schools, and ultimately at all educational levels due to their richness to transmit new knowledge, and make a meaningful learning that generates conceptual changes.

ACKNOWLEDGEMENTS

We appreciate your attention and interest in this topic, which is of great importance, in the process of improving the teaching-learning processes.

## AUTHORS

**Edinson Espitia Correa** was born on April 9, 1988 in Colombia in the department of Córdoba. My educational background is a professional title BACHELOR'S DEGREE IN BASIC EDUCATION WITH EMPHASIS IN COMPUTER TECHNOLOGY at the Caribbean University Corporation close to the date of the degree on December 18, 2014 in Sincelejo Sucre. Postgraduate specialization in ETHICS AND PEDAGOGY at the Juan de Castellanos University Foundation dated June 30 2016. 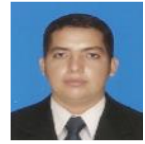
Edinson Espitia Correa with experience in teaching at the Santa Fe Ralito Educational Institution in the position of FULL TIME TEACHER
ES. Espitia

**Pedro Nel Hernández Álvarez** born in coveñas sucre-a on September 4, 1992 in Colombia my educational background professional studies BA in PHYSICAL EDUCATION, RECREATION AND SPORTS at the University of Córdoba graduated on December 23, 2013 in Montería Córdoba. Postgraduate SPECIALIZATION IN PHYSICAL ACTIVITY AND HEALTH graduated on August 02, 2018. 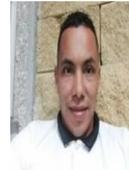
Pedro nel Hernández Álvarez with experience in teaching at the Piñalito Educational Institution Pueblo Nuevo Córdoba in the position of ACTIVE FULL-TIME TEACHER
ES.Hernandez.